\documentclass[pre,twocolumn]{revtex4-1}
\usepackage[dvipdfmx]{graphicx}
\usepackage{bmpsize}
\usepackage{calc}
\usepackage{amsmath}
\usepackage{physics}
\usepackage{bm}          
\usepackage{mathrsfs}  
\usepackage{amssymb}
\usepackage{comment}
\usepackage{cases}
\usepackage{here}
\usepackage{array}
\usepackage[dvipdfmx]{hyperref}
\usepackage{xr}
\usepackage{color}
\usepackage{ulem}
\usepackage{hyperref}

\usepackage{xcolor}

\begin{document}

\title{Emergence of a Giant Rotating Cluster of Fish in Three Dimensions by Local Interactions}
\author{Susumu Ito and Nariya Uchida}
\thanks{uchida@cmpt.phys.tohoku.ac.jp}
\affiliation{Department of Physics, Tohoku University, Sendai, 980-8578, Japan}
\date{\today} 

\begin{abstract}
Schooling fish exhibit giant rotating clusters such as balls, tori, and rings,
among other collective patterns.
In order to account for their giantness and flexible shape change,
we introduce an agent-based model that limits 
the number of agents that each agent can interact with
(interaction capacity).  
Incorporating autonomous control of attractive interactions, 
we reproduce rotating clusters (balls, tori, and rings) that are an order of magnitude larger 
than the interaction range.
We obtained a phase diagram of patterns including polarized schools and swarms.
In our model, the scaling law between 
the number of agents and the projected area of the cluster 
is in good agreement with experimental results.
The model indicates that giant rotating clusters are formed at low interaction capacity,
without long-range interactions or inherent chirality of fish.
\end{abstract}

\maketitle
\section{I\lowercase{ntroduction}}
Cluster formation and collective motion are ubiquitously found in life of various 
organisms~\cite{Conradt2005,Moussaid2009,Vicsek2012}.
Moving clusters are classified into 
``swarms'' of randomly oriented individuals, 
``polarized schools" with directed movement,
and ``vortices'' or rotating clusters~\cite{Delcourt2016}.  
Schooling fish exhibit giant vortices (balls, tori, and rings)~\cite{Simila1997,Parrish2002,Lopez2012,Terayama2015,Masadeh2019}, 
which sometimes contain several thousands of fish
and have a diameter of several ten times the body length~\cite{Terayama2015}. 
Compared to vortices in other biological systems, where 
rotational symmetry is broken by inherent chirality of 
the basic element~\cite{Riedel2005,Sumino2012} 
or by interactions with boundaries~\cite{Wioland2013,Wioland2016},
vortices of fish are unique and highly non-trivial
in that the symmetry is spontaneously broken only by interactions 
between the moving elements~\cite{Tunstrom2013}.

Previous models of fish schools are based on agent-based
approach~\cite{Aoki1982,Reynolds1987,Niwa1994,Vicsek1995}.
Vortices in two and three dimensions are induced by 
various types of interactions:
isotropic attractive and repulsive interactions by
a potential~\cite{DOrsogna2006,Nguyen2012,Chuang2016,Cheng2016},
and asymmetrical interactions via
a viewing angle~\cite{Shimoyama1996,Couzin2002,
Strombom2011,Strombom2015,
Barberis2016,Costanzo2018,Costanzo2019}.
An alternative approach is metric-free models 
that use Voronoi tessellation to determine the neighbors that each agent 
interacts with~\cite{Gautrais2012,Calovi2014,Filella2018}. 
Metric-free interactions are originally introduced for modeling flocks of 
birds~\cite{Ballerini2008,Bode2011,Cavagna2010,Bialek2012} 
as the ``topological interactions'', 
which fix the number of neighbors to interact with. 
It enables a cluster to sensitively react on the motion of a small number of agents,
such as those attacked by a predator, and to flexibly change its shape.
A few experimental studies show that fish also use 
topological interactions~\cite{Faria2010,Herbert2011,Gautrais2012}.
Simulation results with topological interactions are compared to
experimental observations using barred flagtail (\textit{Kuhlia mugil})~\cite{Gautrais2012}.
In two dimensions, 
a model with 
long-range hydrodynamic interactions at low Reynolds number
shows emergence of a vortex which is much larger than characteristic 
length of fish,
and the relation between the number of agents and rotatability~\cite{Filella2018}.
(See Supplemental Material~\cite{Supp1} for detailed discussions of the previous studies.)

In this paper, we propose a new model
that requires only local interactions to form giant vortices in 
three dimensions. 
Here, "giant" means that the cluster size is much larger than the 
interaction range. 
The key idea is to limit the neighbors to interact 
by both their number and distance.
Limiting the maximum number of neighbors that each fish 
can interact with (which we call ``interaction capacity'')
is necessary to avoid the attractive interactions to pile up
and induce unphysically dense clusters.
In fact, there is an experimental evidence that 
attraction is weakened in a cluster of fish~\cite{Katz2011}.
Experiments also show that the attraction and repulsion 
are balanced at a fixed distance for a pair of fish
outside the cluster~\cite{Katz2011, Herbert2011}.
Therefore, we combine the interaction capacity and fixed range of interaction in our model. 
On the other hand, it is also suggested that 
fish use asymmetrical interactions via a blind angle~\cite{Katz2011, Herbert2011}. 
Theoretically, many previous models~\cite{Strombom2011,Barberis2016,Strombom2015,Costanzo2018,Costanzo2019,Couzin2002,Calovi2014} show that introducing the blind angle promotes formation of a rotating cluster.
In the present paper, we 
neglect the blind angle and
show that the interaction capacity is sufficient to reproduce rotating clusters.

We obtained a phase diagram of collective patterns that include
polarized schools, swarms, and 
various rotating clusters (tori, rings, and balls).
In particular, we reproduced a giant ball-shaped rotating cluster
which is similar to a ``bait-ball"~\cite{Lopez2012,Masadeh2019}. 
We investigate 
the cluster size over
a wide range of particle numbers
from a few hundred to tens of thousands, 
and 
reproduced an experimentally observed scaling law 
between the number of agent and the cluster size~\cite{Misund1993}.
Furthermore, we reveal that each agent in the rotating cluster 
performs a random motion in the radial direction,
instead of moving rigidly around a certain radius~\cite{DOrsogna2006}.

The paper is organized as follows. 
In Section~\ref{Section2}, we introduce the model incorporating the topological nature
of interactions. 
In Section~\ref{Section3}, we show the patterns of collective motion 
and a scaling law for the cluster size. 
In Section~\ref{Section4}, we analyze the radial motion of individual agents in a rotating cluster.
We discuss the results in comparison with previous studies in Section~\ref{Section5}.

\section{M\lowercase{odel}}\label{Section2}
The model is based on experimentally observed behaviors of fish. 
First, an attractive interaction is balanced with a repulsive interaction at the equilibrium distance $r_e$ for a few fish,
but is weakened in a cluster (while repulsion within $r_e$ remains).
These experimental results are obtained from data for
golden shiners (\textit{Notemigonus crysoleucas})~\cite{Katz2011}
and mosquitofish (\textit{Gambusia holbrooki})~\cite{Herbert2011}
in shallow tanks.
The mechanism of the weakening is not known, but it is argued 
that the longer-range forces (attraction) are ``more likely to cancel out 
when individuals have neighbors on all sides"~\cite{Katz2011}.
This interpretation is generic and not limited to specific species or environment.

Secondly, 
we assume that fish use topological interactions 
with a maximal number of 
interacting neighbors (interaction capacity). 
Topological interactions of fish are considered in a few previous studies:
A group of three-spined sticklebacks (\textit{Gasterosteus aculeatus} L.),
including a robotic fish as a stimulus, 
have orientational correlation
that depend on the topological distance~\cite{Faria2010}.
In an experiment using mosquitofish (\textit{Gambusia holbrooki})),
it is shown that repulsive interaction acts only on 
the nearest neighbor~\cite{Herbert2011}.
Gautrais {\it et al.}~\cite{Gautrais2012} estimated 
the interaction capacity by comparing a simulation model and 
data for a small group barred flagtail (\textit{Kuhlia mugil})~\cite{Gautrais2012}. 
They obtained the best prediction for the interaction capacity $K=6-8$
while the error in distance was minimal for  $K=3$. 
Although the values of the interaction capacity vary 
depending on the species and the type of interactions, 
these studies equivocally suggest the topological nature of 
the interactions between fish.
In our model, we estimate the interaction capacity as follows:
It is observed for three different species of saltwater fish
, which are cod (\textit{Gadus morhua}), saithe (\textit{Pollachius virens}),
and herring (\textit{Clupea harengus}),
that up to the third nearest neighbors are 
distributed within the order of one body lengths (BL) from each fish~\cite{Partridge1980}.
All of these species form three-dimensional schools but 
in varying degrees:
cod is loosely schooling, herring is strongly schooling, and saith is in the middle.
This distance is close to the value of $r_e$ obtained in Refs.~\cite{Katz2011,Herbert2011}, 
which implies that the interaction capacity of some fish is a few.
Given the situation, we limit the topological interaction within the interaction range $r_e$.

Finally, we take notice of an acceleration mode of fish to escape from predators called ``fast-start"~\cite{Jayne1993,Spierts1999,Wakeling2005}. Some fish in a cluster change their velocity by swimming away from predators, and the change propagates to other fish and causes a dynamic shape change of the cluster~\cite{Hunter1969,Radakov1973,Godin1985,Rosenthal2015}.
Physiologically, fast-start results from the firing of neurons
called Mauthner cells (M-cells)~\cite{Jayne1993,Wakeling2005}.
The firing of M-cells can be triggered
by visual or acoustic stimuli or spontaneously at low rate~\cite{Rosenthal2015}.
Fast-start has duration of about $\tau=0.1$ sec~\cite{Webb1978,Domenici1997}.
We note that
fish recognizes other fish's position by using visual information~\cite{Pitcher1980}.
We model fast-start by turning on attraction when a fish has only a small number of neighbors within $r_e$.
This is based on the following observation:
Rosenthal {\it et al.}~\cite{Rosenthal2015} studied the initiation and 
propagation of evasive motion in a school of golden shiners, 
and showed that the fish that initiates a collective evasive behavior
(``initiator") tends to be located near the cluster boundary.
The fish near the cluster boundary would have 
a smaller number of neighbors than those inside the cluster.
Also, it would have a wider area of view and recognize a predator 
earlier than those inside.
Turning away from the predator coming from outside
naturally results in moving toward the cluster.
Because incorporation of a predator makes the model very complicated,
we modeled the escape response by the attraction toward other fish, 
which is turned on for fish near the cluster boundary. 
We also note that the observation of the evasive 
behavior~\cite{Rosenthal2015} was done in the absence of predators. 
Note that an experiment shows that fish also have orientational interactions~\cite{Calovi2018}. Integrating these properties, we formulate the model as follows.

\begin{figure*}[!t]
\centering
\includegraphics[width=\linewidth,bb=0 0 360 252]{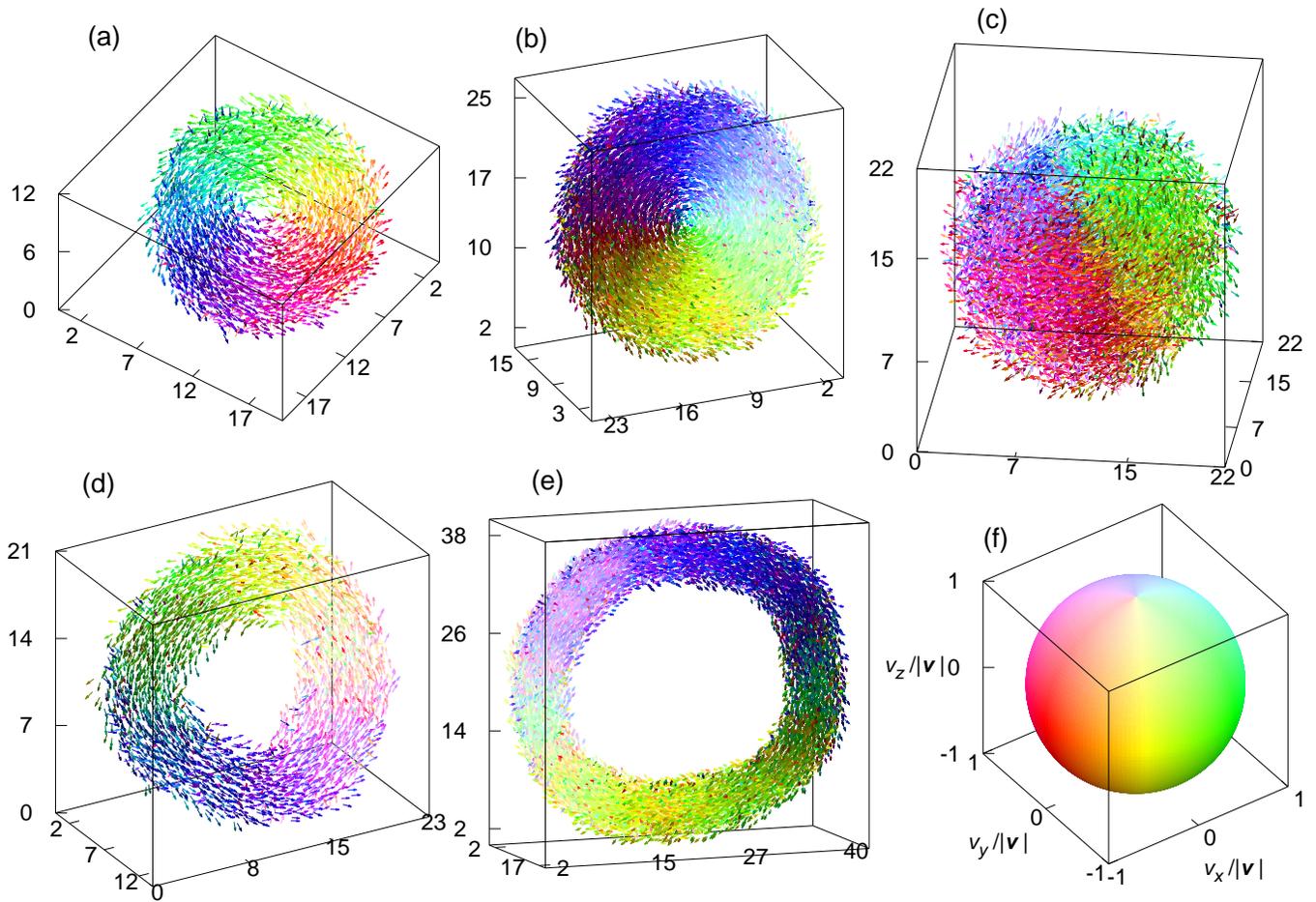}
\caption{(Color online)
Snapshots of fish clusters. Agents are represented by arrows of length 2/3 ($=$1 BL), and the color corresponds to the moving direction of the agent according to the color sphere (f). Shown are only a small part of the simulation box that contains the agents and the origin is shifted for visibility. 
(a) A torus for $N=3000$, $N_u=3$, $\lambda=7.0$. 
(b) A torus for $N=10000$, $N_u=3$, $\lambda=11.0$. 
(c) A rotating ball for $N=10000$, $N_u=1$, $\lambda=11.0$. 
(d) A ring for $N=3000$, $N_u=3$, $\lambda=4.5$. 
(e) A ring for $N=10000$, $N_u=3$, $\lambda=7.0$. 
(f) The color sphere corresponding to agent's direction $\bm{v}/|\bm{v}|=(v_x/|\bm{v}|,v_y/|\bm{v}|,v_z/|\bm{v}|)$. It is a pure hue on the plane of $v_z/|\bm{v}|=0$, and whitish hue when $v_z/|\bm{v}|>0$ and blackish hue when $v_z/|\bm{v}|<0$. See Movie S1-S5 
in Ref.~\cite{SUP_Movies} for dynamics.}
\label{snapshot}
\end{figure*}

\begin{figure*}[!t]
\includegraphics[width=\linewidth,bb=0 0 669 558]{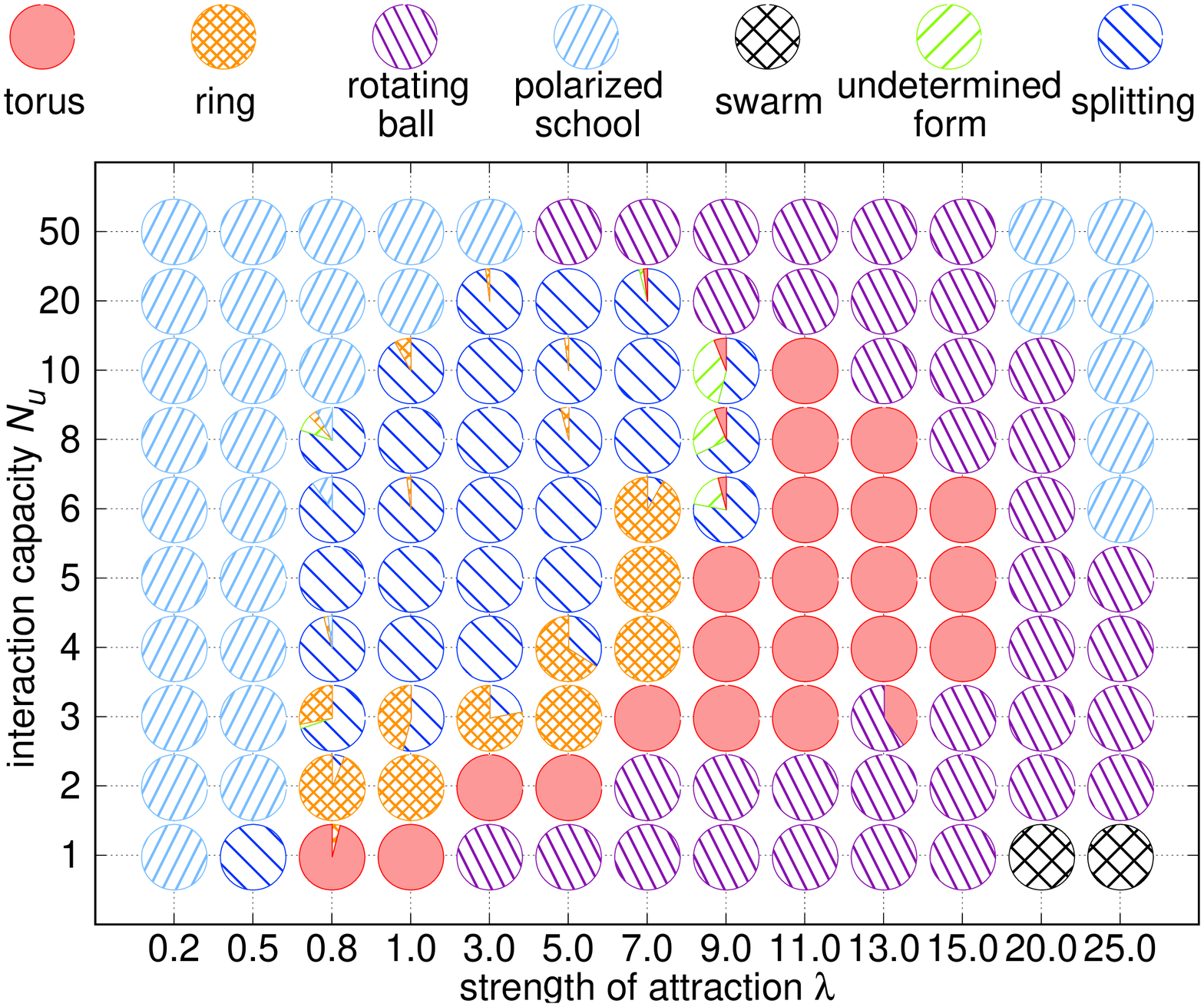}
\caption{(Color online)
Phase diagram of the patterns ($N=3000$) with initial condition (\textit{ii}). The simulation time is up to $t=1500$. The pie chart shows the frequency of occurrence of each pattern in 50 runs performed for each parameter set ($\lambda,N_u$).}
\label{pie chart}
\end{figure*}

We consider $N$ agents moving in a cubic box of size $L$ with the periodic boundary condition. Let $\bm{r}_i$ and $\bm{v}_i=d\bm{r}_i/dt$ ($i=1,2,\ldots, N$) be the position and velocity of the $i$-th agent, respectively. The velocity is set to relax to the sustained speed $v_0$ for an isolated agent, and is modified by orientational, repulsive, and attractive interactions (fast-start) with the neighbor agents. The orientational and repulsive interactions act with up to $N_u$-th nearest neighbors in the radius $r_e$, where $N_u$ is the interaction capacity. The set of agents that can interact with the $i$-th agent by orientational and repulsive interactions is denoted by $\mathcal{L}_i$, and the number of them by $|\mathcal{L}_i|$. The ``occupancy ratio" $\eta_i=|\mathcal{L}_i|/N_u$ is less than or equal to unity and shows how much of the interaction capacity is used. The attractive interaction operates at distances between $r_e$ and $r_a$, and the set of agents that can attract the $i$-th agent is denoted by $\mathcal{A}_{i}=\left\{j\middle|r_e<|\bm{r}_{ij}|\leq r_a\right\}$
with its size $|\mathcal{A}_i|$.
Thus our model combines topological and fixed-range interactions.

The equation of motion is
\begin{eqnarray}
\label{velocity}
\tau_0\dv{\bm{v}_i}{t}&=&(v_0-|\bm{v}_i|)\frac{\bm{v}_i}{|\bm{v}_i|}+\frac{1}{|\mathcal{L}_{i}|}\sum_{j \in \mathcal{L}_{i}}g(|\bm{r}_{ij}|)(\bm{v}_j-\bm{v}_i)\nonumber\\
&&+\frac{1}{|\mathcal{L}_{i}|}\sum_{j \in \mathcal{L}_{i}}g(|\bm{r}_{ij}|)\left(v_r\frac{\bm{r}_{ij}}{|\bm{r}_{ij}|}-\bm{v}_i\right)\nonumber\\
&&+\frac{\Lambda_i(t)}{|\mathcal{A}_{i}|}\sum_{j \in \mathcal{A}_{i}}\left(v_a\frac{\bm{r}_{ji}}{|\bm{r}_{ji}|}-\bm{v}_i\right),
\end{eqnarray}
where $\bm{r}_{ij}=\bm{r}_i-\bm{r}_j$. On the left-hand side, we set the mass of the agent to be unity and introduce the characteristic time-scale $\tau_0$. The four terms on the right-hand side represent the self-driving force and the orientational, repulsive, and attractive interactions, respectively. The constants $v_r$ and $v_a$ are the speed of collision avoidance and the maximum reach speed of fast-start, respectively. The repulsive and orientational interactions are enhanced due to body contact (excluded volume effect), which we incorporate into the function
\begin{align}
\label{exclusion func}
g(|\bm{r}|)= \left\{ \begin{array}{ll}
\frac{r_b}{|\bm{r}|} & [|\bm{r}|\leq r_b], \\
1 & [|\bm{r}|>r_b],
\end{array} \right.
\end{align}
where $r_b$ is the body length.

We define the dimensionless strength of attraction $\Lambda_i(t)$ in order to incorporate the screened attractive force in a cluster and the duration of attraction by fast-start. The function $\Lambda_i(t)$ changes in time depending on the history of the size of $\mathcal{L}_i(t)$ as follows. The attraction is turned on ($\Lambda_i(t)=\lambda$) at the moment when $|\mathcal{L}_i(t)|$ becomes smaller than $N_u$. The attraction lasts for the duration $\tau$. If $|\mathcal{L}_i(t)|=N_u$ after the period $\tau$, the attraction is switched off ($\Lambda_i(t)=0$). Otherwise, if $|\mathcal{L}_i(t)|<N_u$ after the first period, the attraction is maintained for another period $\tau$, and this will be repeated until we finally get $|\mathcal{L}_i(t)|=N_u$; see Fig. S1(a)-(c) for graphical illustration~\cite{Supp2}.

Numerical integration of the equation of motion is carried out by the Runge-Kutta method, and the time step $dt= 0.005$ is used unless otherwise stated. We rescale 
all lengths by the radius of equilibrium $r_e=1.5$ BL~\cite{Katz2011,Herbert2011} 
and time by the characteristic timescale $\tau_0=1$ sec, 
which is estimated from 
the experiments~\cite{Wu1977,Videler1991,Lauder2005,Katopodis2012,Tytell2004,Wise2018}: henceforth $r_e=1$ and $\tau_0=1$. Other parameters are also estimated from 
the experimental values as $r_a=5,v_0=1,v_r=1,v_a=5$, and $\tau=0.1$. The interaction capacity $N_u$ and the strength of attraction $\lambda$ are treated as adjustable parameters of the model. The numerical simulation is carried out with $N=3000$ or $10000$ agents and with two types of initial conditions: (\textit{i}) The agents have randomly distributed positions and directions with the same speed $v_0$ ($L=35, 40$). (\textit{ii}) The agents are randomly distributed in a single spherical cluster with the same direction and speed ($L=120$). 
Details of the parameter values~\cite{Supp2} and the initial conditions~\cite{Supp3} are given in Supplemental Material.

\section{C\lowercase{luster} S\lowercase{hapes, }P\lowercase{hase} D\lowercase{iagram, and }S\lowercase{caling} L\lowercase{aw}}\label{Section3}
Typical snapshots at dynamically steady states ($t=1500$) are shown in Fig. \ref{snapshot}. The initial condition (\textit{i}) is used for (a)-(c) and (\textit{ii}) in (d)-(e). Giant vortices are spontaneously formed by choosing appropriate parameter values. Fig. \ref{snapshot}(a)-(b) display giant ``tori" whose sizes are larger than $r_a$. Shown in Fig. \ref{snapshot}(c) is a ``rotating ball", a spherical vortex looking like a ``bait-ball"~\cite{Lopez2012,Masadeh2019}. Fig. \ref{snapshot}(d)-(e) show ``rings" with holes larger than the interaction range.

To classify the cluster shapes quantitatively, 
we monitored the outer and inner radii of the cluster, the principal moments of inertia, as well as the orientational and rotational order parameters, all of which 
converged by the time $t=1500$ if the cluster is stable 
(see Supplemental Material for their definitions and Fig. S4 for time evolution~\cite{Supp4}). 
The clusters are classified into 7 patterns: (\textit{i}) rotating ball, (\textit{ii}) torus, (\textit{iii}) ring, (\textit{iv}) polarized school, (\textit{v}) swarm, (\textit{vi}) splitting, and (\textit{vii}) undetermined form. (\textit{i}) A rotating ball is a rotating cluster that is almost spherical and hence has small differences between its principal moments of inertia. (\textit{ii}) A torus is a rotating cluster with a hole at its center (the inner radius $<r_e$). (\textit{iii}) A ring is a rotating cluster with the inner radius larger than $r_e$. (\textit{iv}) A polarized school is a cluster in which the agents have almost the same direction, and thus the cluster shows directed movement. (\textit{v}) A swarm is an orientationally disordered cluster in which no characteristic order in the orientation of the agents is observed. (\textit{vi}) Splitting is a state in which the initial cluster is unstable and is divided into several clusters. (\textit{vii}) An undetermined form means a single cluster whose shape is constantly changing and the fluctuation of the order parameters are large. (See Fig. S5 for the quantitative definition of these patterns, and Fig. S3 in Supplementary Material~\cite{Supp4}
and Movies S6, S8 in Ref.~\cite{SUP_Movies} for the shapes of non-rotating clusters.)

\begin{figure}[!t]
\includegraphics[width=\linewidth,bb=0 0 360 252]{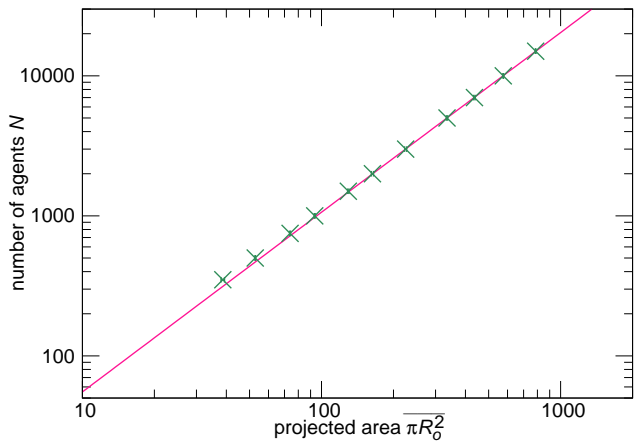}
\caption{(Color online)
Scaling relation between the projected area $\overline{\pi R_o^2}$ and the number of agents $N$ by Log-Log plot ($N_u=3,\lambda=11.0$). The data points (crosses) represent the arithmetic average of the time-averaged $\pi R_o^2(t)$ obtained in each of the 10 simulations, and the error bars represent the standard deviation. 
Fitting by the power law $N \propto \left(\overline{\pi R_o^2}\right)^\nu$ gives $\nu=1.283$ 
(the solid line) over the range $350 \le N \le 15000$.}
\label{scaling law}
\end{figure}

We run the simulation for 50 times for each parameter set ($\lambda,N_u$) with $N=3000$,
and obtained the phase diagram in Fig. \ref{pie chart}. 
It shows the frequency of occurrence of each pattern by pie charts.
Vortex-type clusters (rings, tori, and rotating balls) are obtained in a wide range of parameters, 
although we used the initial conditions in which the agents are aligned. 
The vortex-type clusters frequently appear especially for small $N_u$ and large $\lambda$. 
On the other hand, a polarized school mainly emerges for large $N_u$ and small $\lambda$. 
Splitting and undetermined form are observed at the transition from a polarized school with 
small $\lambda$ to a ring or a torus.

The size of a rotating cluster
changes non-monotonically with $N_u$ with a peak at $N_u=2$ or $3$ (see Fig. S10(a)~\cite{Supp5}),
while the size and period of rotation are decreasing functions of $\lambda$ and increasing functions of $N$ (see Fig. S11~\cite{Supp5}). The average orbital length of each agent 
is proportional to the average period as shown in Fig. S11(c)~\cite{Supp5},
and the proportional coefficient gives the averaged tangential velocity within the cluster $v=0.45$.

In Fig.~\ref{scaling law}, we plot the number of agents $N$ versus the projected area of a cluster 
on the plane perpendicular to the vortex axis
for $N_u=3$, $\lambda=11.0$.  For this parameter set, a rotating cluster 
without a hole is found over a wide range of $N$; see Fig. S13~\cite{Supp5}. 
Therefore, the projected area is estimated by $\overline{\pi R_o^2}$,
where $R_o$ is the outer radius of the cluster and time average is taken.
We find that the size-area relation is well fitted by the scaling law $N \propto \left(\overline{\pi R_o^2}\right)^\nu$ 
with $\nu=1.283 \pm 0.004$.


\begin{figure}[!b]
\includegraphics[width=\linewidth,bb=0 0 360 252]{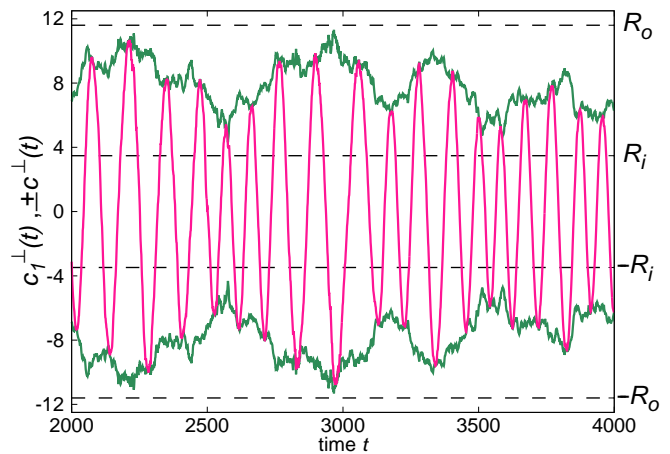}
\caption{(Color online)
Time evolution of the radial distance $c^\perp(t)$ of a certain agent in a ring, and $c_1^\perp(t)=c^\perp(t)\cos\phi(t)$ where $\phi(t)$ is the azimuthal angle in the projected plane ($N=3000,N_u=3,\lambda=4.5$). This orbit is obtained from $t=2000$ to $t=4000$ with the initial condition (\textit{ii}). The upper solid line shows the radial distance $c^\perp(t)$ between the agent and axis of rotation of the cluster, and the lower solid line shows $-c^\perp(t)$. The oscillating curve between the $\pm c^\perp(t)$ lines shows $c_1^\perp(t)$. The dashed lines represent the time averages of the outer radius $\pm R_o(t)$ and the inner radius $\pm R_i(t)$ over the same time period.}
\label{orbit}
\end{figure}

\section{M\lowercase{otion of }I\lowercase{ndividual} A\lowercase{gents}}\label{Section4}
Next we analyze the orbit of each agent in the cluster. The orbit is projected onto a plane perpendicular to the vortex axis, which is parallel to the total angular momentum of the cluster by definition.
Fig. \ref{orbit} shows time-evolution of the radial distance $c^\perp$ of a randomly chosen agent and one of its in-plane coordinate $c_1^\perp = c^{\perp} \cos \phi$. From the plot of $c_1^\perp(t)$, we see that the agent shows a nearly periodical motion with a period of about $100\tau_0$. 
The plot of $c^\perp(t)$ shows that the agent moves back and forth between the outer and inner regions of 
the cluster, and hence the orbit is deviated from a circular trajectory. Although the back-and-forth motion is random, its time-scale is comparable to several orbital periods. 
To characterize the random motion, 
we measured the autocorrelation function $G(\Delta t)$ of the radial distance $c^\perp(t)$, where $\Delta t$ is the time lag. 
We find that $G(\Delta t)$ decays more rapidly than exponentially, 
which means that some disturbance is added to a simple random walk in the radial direction. 
The disturbance is attributed to the motion of the vortex axis, 
and is stronger for a cluster with a smaller aspect ratio (radius/height) and hence for larger $\lambda$,
as demonstrated by Eq. S24 and Fig. S9(a)~\cite{Supp6}.

The number density of agents has a peak at finite radial distance
not only for a torus and a ring (which is trivial by definition) but also for a rotating ball;
the peak moves toward the center as $\lambda$ is increased (see Fig. S14~\cite{Supp7}). 
On the other hand, the occupancy ratio is saturated at $\eta =1$  except at the surface regions 
of width $\sim r_e$, where the agents are attracted
by those inside the cluster (see Fig. S14(a)~\cite{Supp7}). 
We also studied the spatial distribution of the velocity in rotating clusters,
and found that it has peaks at the surfaces while it is constant inside the cluster 
 (see Fig. S15~\cite{Supp7}).
\section{D\lowercase{iscussion}}\label{Section5}
We found that giant rotating clusters (torus, ring, and ball) emerge by reducing the interaction capacity
$N_u$, and without assistance of asymmetrical interactions via a blind angle. 
Our model exhibits, in addition to a torus and a ring that are often found in previous 3D models~\cite{Nguyen2012,Chuang2016,Strombom2015,Couzin2002},
a giant ball-shaped rotating cluster, which is similar to
a ``bait-ball"~\cite{Lopez2012,Masadeh2019}, which is almost spherical for $N_u=1$
(see Fig. S10(b)~\cite{Supp5}).
In some models using isotropic potential~\cite{DOrsogna2006,Cheng2016,Nguyen2012,Chuang2016}, 
there are parameter regions where either schools or mills appear 
depending on the initial conditions. In contrast, in our model, stable tori and rotating balls 
are always found in different parameter regions even though we start from oriented 
clusters (initial condition (\textit{ii})) as shown Fig. \ref{pie chart}.

On the other hand, the cluster size decreases as $N_u$ is increased, and becomes comparable to 
the interaction range for $N_u=50$ (see Fig. S10(a)~\cite{Supp5}). 
Therefore, the interaction is almost global when $N_u=50$.
We obtained polarized schools when $N_u$ is large (see Fig. \ref{pie chart}), 
which is consistent with a previous 3D model without interaction capacity 
that obtained schools 
over wide parameter regions~\cite{Nguyen2012} 
(in  contrast to a 2D model with the same potential~\cite{DOrsogna2006} ).

In our model,  the cluster size is an increasing function of the number of agents 
(see Fig. S11(a) and Fig. S13(b)~\cite{Supp5}).
In contrast, in a previous model with an isotropic potential~\cite{DOrsogna2006},
the cluster size decreases as the number of agents increases.
This is an essential difference arising from  
non-additivity due to interaction capacity in our model;
attractive forces act on only the agents near the surface of a cluster as shown in Fig. S14(a)~\cite{Supp7}.
Furthermore, the rotational order parameter $M$ increases monotonically as the number of agents 
increases (see Fig. S13(a)~\cite{Supp5}), unlike \cite{Calovi2014,Filella2018} where $M$ reaches its peak at several hundred agents.

From here, we compare our results with experimental results.
In the case of $N_u\lesssim3$, we obtain a variety of rotating patterns
in a wide range of $\lambda$ (see Fig.~\ref{pie chart}).
This results support our estimate of $N_u$ (a few).
Furthermore, we can roughly estimate the strength of attraction $\lambda$
from the experimental data:
the acceleration in the fast-start has a peak value around 100 BL s$^{-2}$ \cite{Domenici1997},
and its time-average is estimated to be several tens BL s$^{-2}$
because it has a spike in the time course \cite{Wise2018}.
The time constant of fast-start as the ratio of the velocity $v_a$ to acceleration, therefore,
is on the order of $\mathcal{O}(10^{-2})$ to $\mathcal{O}(10^{-1})$ s.
It corresponds to $\tau_0/\lambda$ in our model,
from which $\lambda$ is estimated to be on the order of $\mathcal{O}(1)$ to $\mathcal{O}(10)$.
This is the range for which we obtained the various rotating clusters.

The actual size of clusters with 3000 sardines is reported in Ref.~\cite{Terayama2015}: 
The inner and outer radii of a torus-type cluster is $1\sim 2$ BL and $13\sim17$ BL, respectively
(We use Fig. 8 (a-b) and (e-f) in Ref.~\cite{Terayama2015} as data for comparison.). These values are close to the radii we obtained 
for $N=3000,N_u=3$, and $\lambda=7.0$, which are $1$ BL (inner radius) and 
$15$ BL (outer radius); see Fig. S11(a)~\cite{Supp5}.
As for the scaling law of the cluster size, experiments report $\log(\mathrm{biomass})=1.329\times\log(\mathrm{school~area})+0.428$ for herring (\textit{Clupea harengus}) and mackerel (\textit{Scomber scombrus})
\cite{Misund1993}.
The exponent $\nu=1.283$ in our model is in good agreement with the experimental value.
The exponent should be $1$ for a disk of constant thickness and density, 
while $\nu=3/2=1.5$ for a sphere of constant density.
The experimental and our numerical results show that 
rotating clusters are intermediate between a disk and a sphere.

The velocity is almost constant inside the cluster (see Fig. S15~\cite{Supp7}), 
while in the experiment, the velocity increases with the distance from the vortex axis~\cite{Terayama2015}.
This could be improved by introducing heterogeneity of swimming velocity,
which makes faster agents distributed in the outer side of a cluster~\cite{Costanzo2019}. 
Heterogeneity in other characteristics, such as the blind angle~\cite{Romey2013},
also controls the shape and size of the cluster, and might affect the velocity distribution.

Finally, 
it is important to reproduce the verticality of the vortex axis in actual fish clusters,
which is presumably due to gravity and upward movement of predators 
near water surface \cite{Simila1997,Masadeh2019}. 
Inclusion of these effects into the model will be an interesting issue for the future.


\end{document}